\documentstyle[aps,pra]{revtex}

\begin{document}
\title{Grover's Quantum Search Algorithm \\
for an Arbitrary Initial Amplitude Distribution
}
\author{Eli Biham$^1$, Ofer Biham$^2$, David Biron$^2$, Markus
Grassl$^3$ and Daniel A. Lidar$^4$}
\address{
$^1$Computer Science Department, Technion, Haifa 32000, Israel \\
$^2$Racah Institute of Physics, The Hebrew University, Jerusalem
91904, Israel\\ 
$^3$Institut f\"{u}r Algorithmen und Kognitive Systeme,
Universit\"{a}t Karlsruhe, 
\hbox{Am Fasanengarten 5}, D--76128 Karlsruhe, Germany\\
$^4$Department of Chemistry, University of California, Berkeley, CA
94720, USA}

\maketitle

\begin{abstract}
Grover's algorithm for quantum searching is generalized
to deal with arbitrary initial complex amplitude
distributions. First order linear 
difference equations are found for
the time evolution of the amplitudes of the marked and 
unmarked states. These equations are solved exactly. 
New expressions are derived for the optimal time of measurement and the
maximal probability of success.
They are found to depend on the averages
and variances of the initial amplitude distributions 
of the marked and unmarked states,
but not on higher moments. Our results imply that Grover's algorithm
is robust against modest noise in the amplitude initialization procedure. 
\end{abstract}


It is now firmly established that there exists a gap between the
computational power of quantum and classical computers. A
dramatic example of the speed-up offered by quantum computers is
Grover's quantum search algorithm \cite{Grover96,Grover97} 
for finding a marked element
among $N$ possible input values, in the presence of an oracle. 
On average a classical computer
would need $N/2$ oracle-queries, whereas a quantum computer can
accomplish the same task using merely $O(\sqrt{N})$ queries. The
importance of Grover's 
result stems from the fact that it proves the enhanced power of
quantum computers compared to classical ones for a whole class of
oracle-based problems, for which the bound 
on the efficiency of classical
algorithms is known. 

Grover's algorithm can be represented as searching a
preimage of an oracle-computable boolean function, 
which can only be computed
forward, but whose inverse cannot be directly computed.
Such a function is $F:  D  \rightarrow \{0,1\}$
where $D$ is a set of $N$ domain values (or states)
and the preimages of the value 1 are called the {\em marked} states.
The problem is to identify one of the marked states, i.e., some
$v \in D$ such that $F(v)=1$.
Problems of this type are very common.
One important example, from cryptography,
is searching for the key $K$ of the Data Encryption Standard
(DES) \cite{Stinson95}, given a known plaintext $P$ 
and its ciphertext $C$,
where $F=1$ if the pair of plaintext and ciphertext match [i.e.,
$E_K(P)=C$ where $E_K$ is the encryption function] and $F=0$ otherwise.
Other examples are solutions of NP and NP-complete problems, which
include virtually all the difficult computing problems in practice
\cite{Garey79}.

A large number of results followed Grover's discovery.
These results include a
proof \cite{Zalka97} that the algorithm is as efficient as theoretically
possible \cite{Bennett97}; a variety of applications in which the
algorithm is used in the solution of other
problems
\cite{Durr96,Grover97b,Grover97c,Brassard97a,Terhal97,Brassard98,Cerf98,Farhi98};
and recently, an experimental implementation using a
nuclear magnetic resonance 
(NMR) quantum
computer \cite{Chuang98}. 
Several generalizations of Grover's original
algorithm have been published, the first of which dealt with the case
of more than one marked state \cite{Boyer96}. 
The algorithm was further generalized 
by allowing an arbitrary (but constant) unitary
transformation to take the place of the Hadamard transform in the
original setting \cite{Grover98a}. 

In this Rapid Communication, we generalize Grover's algorithm by allowing for an 
{\em arbitrary complex initial amplitude distribution}. 
We present an exact solution for the time evolution of the
amplitudes under these general initial conditions. 
We find that the generalized search algorithm still requires
$O(\sqrt{N/r})$ iterations, 
where $r$ is the number of marked states,
although the maximal success probability can be
small for certain unfavorable initial amplitude distributions. The
case of an arbitrary initial amplitude distribution is particularly
relevant in the presence of unitary errors in the gates implementing the
initialization step, such as over- or under-rotations. Such errors can
result in a deviation from the 
uniform initial amplitude distribution assumed in the usual treatment
of Grover's algorithm, and as detailed below, our analysis shows that
the algorithm will still work in the presence of modest errors.

We will now present the modified Grover algorithm and derive difference 
equations for the time evolution of the amplitudes in it. 
We then solve these equations exactly and analyze the results.
Let $k(t)$ [$l(t)$] 
denote the amplitude of the marked [unmarked] states 
after $t$ iterations of the algorithm.
It was shown in \cite{Boyer96}
that the amplitude of the marked 
states increases as: $k(t)=\sin [\omega (t+1/2)]/\sqrt{r}$, where
$\omega =2\arcsin (\sqrt{r/N})$. At the same time the amplitude of the
unmarked states decreases as: 
$l(t)=\cos [\omega (t+1/2)]/\sqrt{ N-r}$. 
For $N \gg r$ the
optimal time to measure and complete the calculation is after
$T=O(\sqrt{N/r})$ iterations, when $k(t)$ is maximal. In our modified
algorithm we simply omit the initialization step from Grover's
original algorithm. It thus consists of the following
stages: 

\begin{enumerate}
\item Use any initial distribution of marked and unmarked states,
e.g., the final state of any other quantum algorithm (do {\it not}
initialize the system to the uniform distribution).

\item  Repeat the following steps $T$ times:

\begin{description}
\item[A.]  Rotate the marked states by a phase of $\pi $ radians.

\item[B.]  Rotate all states by $\pi $ radians around the average
amplitude of {\it all} states. This is done by (i) Hadamard
transforming every qubit; (ii) rotating the $|0\rangle$ state by a
phase of $\pi $ radians; (iii) again Hadamard transforming every
qubit. 

\end{description}

\item  Measure the resulting state.
\end{enumerate}

Next, we analyze the time evolution of the amplitudes in the
modified algorithm with a total of $N$ states. Let the marked
amplitudes at time $t$ be denoted by $k_{i}(t)$, $i=1,\dots ,r$ and
the unmarked amplitudes by $ l_{i}(t)$, $i=r+1,\dots ,N$, where the
initial distribution at $t=0$ is arbitrary. Without loss of generality
we assume that the number of marked states satisfies $1\leq r\leq
N/2$. Let the averages of the amplitudes be denoted by

$$
\bar{k}(t)=\frac{1}{r} \sum_{i=1}^{r}k_{i}(t) 
$$

\noindent
for the marked states, and by

$$
\bar{l}(t)=\frac{1}{N-r} \sum_{i=r+1}^{N}l_{i}(t) 
$$

\noindent
for the unmarked states.
The key observation is that the entire dynamics dictated by Grover's
algorithm can be described in full by the time-dependence of the {\it 
averages}.
Let us define

\begin{equation}
C(t) = \frac{2}{N}\left[ 
(N-r)\bar{l}(t)-r\bar{k}(t)\right].   
\label{eq:C}
\end{equation}

\noindent
Consider any marked state
$k_{i}(t)$. 
In each step of the algorithm
this state is flipped to 
$k_{i}^{\prime }(t)=-k_{i}(t)$,
so that the marked average becomes 
$\bar{k}^{\prime }(t)=-\bar{k}(t)$. 
The unmarked states, on the other hand, 
do not flip, so that the average over 
{\it all} states after the flip
is: 
$x(t)= \frac{1}{N}[r\,\bar{k}^{\prime}(t)+
(N-r)\,\bar{l}(t)]=C(t)/2$. 
Rotation by $\pi$ radians around the average 
is by definition: 
$k_{i}^{\prime }(t)\rightarrow 2x(t)-k_{i}^{\prime }(t)$
and 
$l_{i}(t) \rightarrow 2x(t)-l_{i}(t)$.  
Hence,
$k_{i}(t) \rightarrow C(t)+k_{i}(t)$ 
and 
$l_{i}(t) \rightarrow C(t)-l_{i}(t)$. 
Therefore,
the time evolution of all amplitudes 
(of both marked and unmarked states) is 
independent of the state index, and satisfies: 
\begin{eqnarray}
k_{i}(t+1) &=&C(t)+k_{i}(t)\qquad\qquad i=1,\dots ,r
\label{eq:recur_k} \\
l_{i}(t+1)
&=&\rlap{$C(t)-l_{i}(t)$}\phantom{C(t)+k_{i}(t)}\qquad\qquad
i=r+1,\dots ,N. 
\label{eq:recur_l}
\end{eqnarray}

\noindent
By averaging over the states in Eqs. 
(\ref{eq:recur_k})
and
(\ref{eq:recur_l})
we find that the average marked and unmarked amplitudes 
obey first order linear coupled difference equations:

\begin{eqnarray}
\bar{k}(t+1) &=&C(t)+\bar{k}(t)         \label{eq:k-ave} \\
\bar{l}(t+1) &=&C(t)-\bar{l}(t){\rm .}  \label{eq:l-ave}
\end{eqnarray}

\noindent
These equations can be solved for 
$\bar{k}(t)$ 
and 
$\bar{l}(t)$, 
and along with the initial distribution this 
yields the exact solution for
the dynamics of all amplitudes.
We proceed to solve the recursion formulae for arbitrary complex
initial conditions. 
Let:

\begin{eqnarray*}
f_{+}(t) &=&\bar{l}(t) + i\sqrt{\frac{r}{N-r}}\bar{k}(t)       \\
f_{-}(t) &=&\bar{l}(t) - i\sqrt{\frac{r}{N-r}}\bar{k}(t){\rm .}
\end{eqnarray*}

\noindent
Using the recursion formulae
(\ref{eq:k-ave}) 
and
(\ref{eq:l-ave}) 
and
a few steps of algebra employing
the definition of
$C(t)$ given in
(\ref{eq:C}), 
we find that
$f_{+}(t+1) = e^{i\omega} f_{+}(t)$
and
$f_{-}(t+1) = e^{-i\omega} f_{-}(t)$.
Here $\omega$, 
which is real and satisfies

\begin{equation}
\cos\omega= 1 - 2 {r \over N},
\label{eq:omega} 
\end{equation}

\noindent
is identical to the frequency found by Boyer {\it et al.} in
\cite{Boyer96}.
The time evolution can now be written as
\begin{eqnarray*}
f_{+}(t) &=& e^{i\omega t}f_{+}(0) \\ 
f_{-}(t) &=& e^{-i\omega t}f_{-}(0). 
\end{eqnarray*}

\noindent
Clearly,
$|f_{+}(t)|$ 
and
$|f_{-}(t)|$ 
are time independent quantities.
The average amplitudes are

\begin{eqnarray}
\bar{k}(t) &=& -i \sqrt{\frac{N-r}{4r}} 
\left[ e^{i\omega t} f_{+}(0) - e^{-i\omega t} f_{-}(0)\right]  
\label{eq:sol_k} \\
\bar{l}(t) &=& {1 \over 2} 
\left[e^{i\omega t} f_{+}(0) + e^{-i\omega t} f_{-}(0)\right]. 
\label{eq:sol_l}
\end{eqnarray}

\noindent
Together with Eqs. 
(\ref{eq:C})-(\ref{eq:recur_l}) 
this
provides the complete exact solution to the dynamics of the amplitudes
in the generalized Grover algorithm, for arbitrary initial conditions.

We turn to an analysis of several properties of the amplitudes
and to a simplification of the result describing the dynamics.
Let 
$\alpha$ 
and 
$\phi$ 
(real or complex) be chosen such that
$\alpha = \sqrt{f_{+}(0)f_{-}(0)}$, and
$e^{2i\phi}=f_{+}(0)/f_{-}(0)$.
Using Eqs. 
(\ref{eq:sol_k})
and
(\ref{eq:sol_l}),
the average amplitudes can be expressed concisely as follows

\begin{eqnarray}
\bar{k}(t) &=& 	\sqrt{{N-r} \over r} {\alpha} \sin (\omega t+\phi ) 
\label{eq:ksin} \\
\bar{l}(t) &=& \alpha \cos (\omega t+\phi ). 
\label{eq:lcos}
\end{eqnarray}

\noindent 
This shows that there is a $\pi /2$ phase difference between
the marked and unmarked amplitudes: when the average marked amplitude
is maximal, the average unmarked amplitude is minimal, 
and {\it vice versa}.
[Note that when the ratio
$\bar{l}(0)/\bar{k}(0)$ 
is real, 
$\alpha$ 
and
$\phi$ become real, 
with 
$\alpha^2 = |\bar{l}(0)|^2 + |\bar{k}(0)|^2 r/(N-r)$ 
and
$\tan \phi = \sqrt{{r}/{(N-r)}} {\bar{k}(0)}/{\bar{l}(0)}$].
Subtracting Eq.
(\ref{eq:k-ave}) 
from Eq.~
(\ref{eq:recur_k}), 
and Eq.~ 
(\ref{eq:l-ave}) 
from Eq.~
(\ref{eq:recur_l}) 
one finds:

\begin{eqnarray*}
k_i(t+1)-\bar{k}(t+1) &=& k_i(t)-\bar{k}(t) \\
l_i(t+1)-\bar{l}(t+1) &=& -[l_i(t)-\bar{l}(t)] {\rm .}
\end{eqnarray*}

\noindent 
This means that:

\begin{eqnarray}
\Delta k_i &\equiv & k_i(0)-\bar{k}(0) \label{eq:deltak} \\
\Delta l_i &\equiv &  l_i(0)-\bar{l}(0) \label{eq:deltal} 
\end{eqnarray}

\noindent 
are {\it constants of motion}.
This allows us to simplify the expression for the 
time dependence of the
amplitudes:

\begin{eqnarray}
k_i(t) &=& \bar{k}(t) + \Delta k_i \label{eq:ki}\\
l_i(t) &=& \bar{l}(t) + (-1)^t \Delta l_i \label{eq:li}{\rm ,}
\end{eqnarray}

\noindent 
where 
$\Delta k_i$ 
and 
$\Delta l_i$ 
are given by the initial amplitude distribution (at $t=0$).

Eqs.
(\ref{eq:ksin}),
(\ref{eq:lcos}),
(\ref{eq:ki}) and
(\ref{eq:li})
are thus an alternative, simplified form describing the dynamics of 
the amplitudes.
In this picture all marked states evolve in unison so it is sufficient to
follow the time evolution of their average.
The only feature distinguishing the states from one another
is their initial deviation from the average.
The same holds true for the unmarked states, up to an alternation
about their average. 

From Eqs.
(\ref{eq:ki})
and
(\ref{eq:li})
it follows immediately that the variances 

\begin{eqnarray}
\sigma_{k}^{2}(t)&=&\frac{1}{r} \sum_{i=1}^{r}|k_{i}(t)-\bar{k}(t)|^{2}
\label{eq:sigma_k} \\
\sigma _{l}^{2}(t)&=&\frac{1}{N-r} 
\sum_{i=r+1}^{N}|l_{i}(t)-\bar{l} (t)|^{2}
\label{eq:sigma_l}
\end{eqnarray}

\noindent
are time-independent.
Now, when a measurement is performed at time $t$,
the probability that a marked state will be obtained is
$P(t)=\sum_{i=1}^{r}|k_{i}(t)|^{2}$. 
Since all the operators used are unitary, the
amplitudes satisfy the normalization condition:

$$
\sum_{i=1}^{r}|k_{i}(t)|^{2}+\sum_{i=r+1}^{N}|l_{i}(t)|^{2}=1
$$

\noindent
at all times. 
Using
$\overline{(y-\bar{y})^{2}}=\overline{y^{2}}-\bar{y}^{2}$ 
($y$ is a random variable), 
we find 
from Eqs.
(\ref{eq:sigma_k})
and
(\ref{eq:sigma_l}):

\begin{eqnarray*}
\sum_{i=1}^{r}|k_{i}(t)|^{2}&=&r\sigma_{k}^{2}+r|\bar{k}(t)|^{2} \\
\sum_{i=r+1}^{N}|l_{i}(t)|^{2}&=&(N-r)\sigma_{l}^{2}+(N-r)|\bar{l}(t)|^{2}.
\end{eqnarray*}

\noindent
Therefore, the probability of measuring a marked state at time $t$ is given by

\begin{equation}
P(t) = P_{av} - \Delta P \cos 2[\omega t + {\rm Re} (\phi)]
\label{eq:Pt}
\end{equation}

\noindent
where

\begin{eqnarray*}
P_{av} &=& 1 - (N-r) \sigma_l^2 - 
{1 \over 2} \left[ (N-r) \left|\bar l(0)\right|^2 + r \left|\bar k(0)\right|^2 \right] \\
\Delta P &=& \frac{1}{2} \left|(N-r) {\bar l(0)}^2 + r {\bar k(0)}^2  \right|. 
\end{eqnarray*}

\noindent
The maximal value that this probability can obtain during the evolution of the 
algorithm is
\[
P_{\max }= P_{av} + \Delta P.
\]
Given an arbitrary initial distributions of 
$r$ marked and $N\!-\!r$
unmarked states, 
with known averages 
$\bar{k}(0)$ 
and 
$\bar{l}(0)$
respectively, 
the optimal measurement times are after
$$
T= [(j+1/2) \pi - {\rm Re} (\phi)]/\omega
$$
iterations, for
$j=0,1,2,...$  
when the probability of obtaining a marked state is $P_{\max}$.
An important conclusion is that to determine the optimal measurement
times, 
all one needs to know are the average initial amplitudes and
the number of marked states.
Expanding $\omega$ in the expression for $T$
in 
$r/N\ll 1$ 
(at $j=0$) 
one finds that 
the number of iterations before the optimal measurement
probability $P_{\max }$ is obtained is 
$O(\sqrt{N/r})$.
However, the value of  $P_{\max}$ can vary significantly, 
depending on the statistical properties (average and variance) of the
initial amplitude distribution.
The expected number of repetitions of the entire algorithm
until a marked state is obtained
is
$1/P_{\max }$.

We next consider the shapes generated in the complex plane
during the time evolution of 
$\bar{k}(t) = \{ {\rm Re} [\bar{k}(t)], {\rm Im} [\bar{k}(t)] \}$ 
and 
$\bar{l}(t) = \{ {\rm Re} [\bar{l}(t)], {\rm Im} [\bar{l}(t)] \}$. 
Eqs.
(\ref{eq:ksin}) 
and
(\ref{eq:lcos}) 
turn out to be identical to the equations that describe the 
polarization of electromagnetic
plane waves
\cite{Born59}.
By this analogy,
the contours generated by these equations
are ellipses in the complex plane. 
The major axis of the ellipse of
$\bar{l}(t)$ 
subtends an angle $\eta$ with the real axis,
where $\eta$ is given by
$e^{i\eta}=\alpha/|\alpha|$.
The length of the major [minor] axis of the ellipse is 
$a = |\alpha| \cosh ({\rm Im} \phi$)
[$b = |\alpha| \sinh ({\rm Im} \phi$)].
Here $\alpha$ and $\phi$ are the parameters which appear
in Eqs.
(\ref{eq:ksin}) 
and
(\ref{eq:lcos}). 
The ellipse of $\bar{k}(t)$ has a similar shape, but its major axis
subtends an angle $\eta + \pi/2$ with the real axis and its major and
minor axes are longer by a factor of $\sqrt{(N-r)/r}$.

When the ratio 
$\bar{l}(0)/\bar{k}(0)$ 
is real,
one can easily show that
$|f_{+}(0)|=|f_{-}(0)|$.
In this case the amplitudes evolve along a straight line in the
complex plane, in analogy to the case of linear polarization of light,
and
$P_{\max }= 1-(N-r)\sigma_l^2$.
The best case, in which
$P_{\max }= 1$, 
is obtained for Grover's original (uniform amplitudes)
initialization, where 
$\sigma^2_l=0$.

A limit in which the algorithm is totally useless is obtained when
either 
$f_{+}(0)=0$ 
or 
$f_{-}(0)=0$.
In this case the success
probability $P(t)$ remains constant during the evolution of the algorithm.
This corresponds to the case of circular polarization of light.
The worst case appears when $P_{\max}=P(t)=0$ for any $t$.
In this case, which is obtained when
$\sigma^2_k = f_{+}(0) = f_{-}(0) = \bar{k}(0) = \bar{l}(0) = 0$
and 
$(N-r)\sigma_l^2=1$,
the algorithm can never find the marked states.

Finally, consider the case where
the average and variance of the 
initial amplitude distribution 
are {\it not} known, 
but different runs of the algorithm use
initial amplitudes drawn from the same distribution.
Naively, one could pick a random number of iterations
$T_r$ and thus find a marked state with probability
$P(T_r)$.
Correspondingly, the expected number of repetitions of
the entire algorithm using the same $T_r$
would be
$1/P(T_r)$
until a marked state is found.
However, $P(T_r)$ could be very small.
A better strategy is
now shown.
From
Eqs. 
(\ref{eq:omega})
and
(\ref{eq:Pt})
it follows that the period of oscillation of $P(t)$
depends only on $r/N$, while the details of the initial amplitude distribution
are all in the phase $\phi$.
Consider the case where one runs the algorithm twice, 
taking measurements at times 
$T_1$ and $T_2$ 
respectively,
where
$T_2-T_1 = \pi/(2 \omega)$.
From Eq.~(\ref{eq:Pt})
it is clear that in one of the two measurements 
$P(T) \ge P_{av} \ge P_{max}/2$.
In this case we need twice as many repetitions to obtain at least
half the success probability compared to the case when the optimal 
measurement time is known. The slowdown is thus at most a
factor of four.  

In this work we generalized Grover's quantum search algorithm to apply
for initial input distributions which are non-uniform. In fact, it was
shown that by simply omitting the first step of Grover's original
algorithm, wherein a uniform superposition is created over all
elements, a more general algorithm results which
applies to {\it arbitrary} initial distributions. To analyze the
algorithm, we found that the time evolution of the amplitudes of the
marked and unmarked states can be described by first-order linear
difference equations with some special properties. The most important
of these is that all amplitudes essentially evolve uniformly, with the
dynamics being determined completely by the average amplitudes. This
observation allowed us to find an exact solution for the
time-evolution of the amplitudes. A significant conclusion from this
solution is that generically the generalized algorithm also has an
$O(\sqrt{ N/r})$ running time, thus being more powerful than any
classical algorithm designed to solve the same task. An important
future application of these results is in the study of the robustness
of Grover's algorithm against errors in the unitary operations used to
implement the algorithm. Our results imply that the algorithm can
tolerate a moderate amount of noise in the amplitude initialization
procedure. Work extending these results to the case of errors in the
inversion about average step, and in the case of an arbitrary unitary
transformation, is in progress.

This work was initiated during the Elsag-Bailey --
I.S.I. Foundation research meeting on quantum computation in 1997, 
and was completed during the ensuing meeting in 1998.

\end{document}